\documentclass{elsart}
\usepackage{psfig}
\def\spose#1{\hbox to 0pt{#1\hss}}
\def\simlt{\mathrel{\spose{\lower 3pt\hbox{$\mathchar"218$}}
     \raise 2.0pt\hbox{$\mathchar"13C$}}}
\def\simgt{\mathrel{\spose{\lower 3pt\hbox{$\mathchar"218$}}
     \raise 2.0pt\hbox{$\mathchar"13E$}}}

\def\aj{AJ}                   
\def\apj{ApJ}                 
\def\aap{A\&A}                
\def\mnras{MNRAS}             
\def\nat{Nature}              

\begin{document}
\runauthor{R\"ottgering et al. }
\begin{frontmatter}
\title{Proto-clusters and the clustering of distant galaxies and radio sources}
\author[leiden]{Huub R\"ottgering,}
\author[eso]{Emanuele Daddi,}
\author[leiden]{Roderik Overzier,}
\author[durham]{Richard Wilman}

\address[leiden]{Sterrewacht Leiden, PO Box 9513, 2300 RA Leiden, The Netherlands}
\address[eso]{European Southern Observatory, Garching, M\"unich, Germany}
\address[durham]{Department of physics, Durham, United Kingdom}


\begin{abstract}
The clustering properties of objects in 3 different radio surveys
(NVSS, FIRST and BOOTES-WSRT) and 2 near-infrared surveys (the ``Daddi
field'' and the FIRES survey) are investigated and compared with
studies of various samples of galaxies, AGN and clusters.  At $z\sim
1$, it seems that the 2dF optical quasars have a correlation length a
factor of about 2 less than powerful radio galaxies at similar
redshifts. This indicates that these two classes of object can not be
``unified'' by postulating that their main difference is due to their
evolution being at a different stage. It seems much more likely that
these QSOs are predominantly located in field galaxies, 
while the powerful radio sources
are located preferentially in early types. Furthermore, it appears
that both the extremely red objects (EROs) from the IR surveys and the
more luminous radio sources are similarly clustered and as such are
the most clustered objects known in the $z>1$ Universe. Even at $z\sim
3$ the red J-K galaxies from the FIRES survey are similarly strongly
clustered, at a level of about 3 times higher than Lyman break
galaxies.  These clustering properties are consistent with EROs and
radio galaxies being similar objects at different stages of their
evolution. Locally, the most clustered population of objects are
clusters of galaxies. Since the progenitors of these objects  --
proto-clusters -- will therefore also be highly clustered, a good way
to locate proto-clusters is to target fields with very powerful and
potentially highly clustered distant $z>2$ radio sources.  The
techniques that are currently being used for locating and studying
these proto-clusters are briefly discussed.
\end{abstract}

\end{frontmatter}

\section{Introduction}

In the local Universe, the distribution of galaxies is observed to be
complex; massive clusters and super-clusters exist as well as large
empty regions, the so called voids. The clusters seem to be connected
by a web of filaments and walls of galaxies.  For a nice visual
illustration of the beauty of the local galaxy distribution, see for
example the maps based on the two-degree Field (2dF) Galaxy Redshift
survey by Peacock et al. (2001). \nocite{pea01} The spatial fluctuations
in the galaxy distribution can be connected with measurements of the
cosmic microwave background (CMB) anisotropies. Lahav et al. (2002)
\nocite{lah02} showed that within a model based on a flat $\Lambda$CDM
Universe, both measurement sets can be fit simultaneously. Ingredients
of such a model include the growth of clustering of dark matter halos,
evolution of the bias parameter (i.e. how do galaxies trace mass at
different epochs) and the history of merging of galaxies. The present
challenge for observers is to measure the clustering properties not
only for large samples of local galaxies, but also for rarer objects
such as active galaxies. Furthermore, measurements of the clustering
of galaxies, clusters and AGN as a function of redshift, luminosity
and mass will in detail constrain models of the evolution of structure
in the Universe.

In this contribution, we first discuss a number of radio and
near-infrared surveys and the analysis of the clustering of the
various classes of objects in these surveys.  It appeared that radio
galaxies at $z\simgt 1$ are among the most clustered objects known in
the $z \simgt 1$ Universe. They are clustered at a level similar to
extremely red galaxies.  Since in the local Universe clusters of
galaxies are the most clustered objects  (Bahcall and Soneira 1983),
\nocite{bah83} the highly clustered distant radio galaxies seemed good
targets to hunt for forming clusters -- proto-clusters.  Finally, we
briefly discuss the use of powerful radio galaxies as probes of such
distant clusters. For more extensive discussions, we refer to other
contributors to these proceedings, including those of Best, Buttery,
Brand, Croft, Kurk, Wold, and Venemans.

\section{Clustering}

A measure of the clustering of an ensemble of objects is the
correlation function $\xi(r)$, where $r$ is the distance. The
correlation function measures the excess chance over a random
distribution of detecting an object a distance $r + \delta r$ from a
given object. The total chance $\delta P$ of detecting an object
within a volume $\delta V$ is therefore given as:
$$ \delta P = n [ 1+ \xi(r)] \delta V, $$
where $n$ is the volume density of the objects. In most cases, the spatial
correlation function can be parametrised as a power law:
$$\xi(r) = (r/r_0)^{-\gamma},$$ where $r_0$ is the scale length at
which the expected overdensity of objects is twice that over
Poissonian.

Often, for a survey only the positions of objects on the sky are known
and not the true spatial distances.  The angular spatial correlation
$w(\theta)$ is then used to measure the angular clustering:
$$ \delta P = n [ 1+ w(\theta)] \delta \Omega, $$ where $\delta
\Omega$ is an infinitesimal area on the sky at a distance $\theta$
from the given object. The corresponding power law is then defined as:
$$ w(\theta) = A \theta^{1-\gamma}.$$
If for a sample the redshift distribution and amplitude and slope
of  the angular two-point correlation function are known, the
spatial correlation function can be derived, using the Limber
equation. For a detailed account of the usage of the two-point
correlation function we refer to the book by Peebles (1980).
\nocite{pee80}

\subsection{Samples}

During the last few years we have analyzed the clustering
properties in 5 different surveys, the NVSS, FIRST and BOOTES
radio surveys and a wide and a very deep near IR survey. We will
discuss these in turn.

\subsubsection{First and NVSS}

\begin{figure}[ht]
\centerline{\psfig{figure=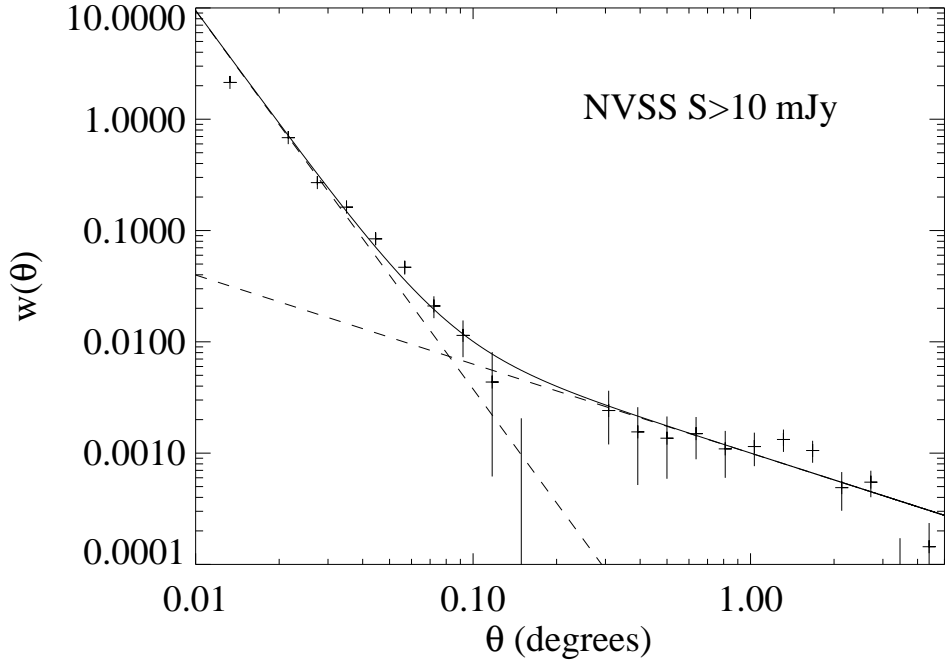,width=0.9\textwidth,clip=}}
\caption{\label{clust} The angular two-point correlation function
  of $S>10$ mJy NVSS sources (from Overzier et al. 2003). }
\end{figure}

The NVSS (Condon et al. 1998) \nocite{con98} and the FIRST (Becker
et al. 1995) \nocite{bec95} radio surveys have both been carried
out by the VLA at 1.4 GHz. They cover large areas of down to a 5
sigma level of 2.5 and 1 mJy respectively.  Overzier et al. (2003)
\nocite{ove03} analysed in detail the clustering properties of
these samples. A similar analysis has been performed by Blake and
Wall (2002). \nocite{bla02} The main result is that the two-point
correlation function can be described as consisting of two power
laws (see Fig. \ref{clust}).

The power law that dominates on the smaller scales is due to
extragalactic radio sources that are often comprised of two distinct
lobes. The median separations of these lobes are 10 - 20 arcsec at
flux levels of a few Jy at 151 MHz, with a long tail to many
arcminutes (Eales et al. 1985). \nocite{eal85c} To model this
power-law, the local angular size distribution as found by Neeser et
al. (1995) \nocite{nee95} was taken and evolved to higher redshift
according the median of the distribution getting smaller at higher
redshifts as $(1+z)^{-1.7}$. Using the radio luminosity function by
Dunlop and Peacock (1990), \nocite{dun90} we can then model the
contribution to the physical doubles to the angular two point
correlation function. This simple model provides a good fit to the
power law on small scales. The power law at larger scales is due to
the cosmological clustering of radio sources.  Interestingly, the
number of radio sources in these surveys is so large that we can
investigate whether the amplitude of the cosmological 
angular correlation function
varies with flux density. Indeed it does, and in the range 3 to 50 mJy
has a fairly constant amplitude of $\sim 10^{-3}$, rising to values of
5 to 10 times higher between 50 and 300 mJy. To convert these
amplitudes into correlation lengths we use redshift distributions as
calculated from the radio luminosity models of Dunlop and Peacock. At
the higher flux density levels these are accurate enough for our
purpose. At the lower flux levels, these models and the resulting
redshift distribution are less accurate, mainly due to a lack of
samples at these flux levels with complete redshift information. One
of the aims of the CENSORS project (Best et al., in prep) is to
provide such a sample. Using these redshift distributions, we find
that the correlation length varies from 5 Mpc for the samples with
flux densities less than 50 mJy to around 15 Mpc at 200 mJy. We note
that if the redshift distribution for the fainter sample turns out to
be broader than inferred from the Dunlop and Peacock model the
resulting correlation length will be larger.

\subsection{Bootes survey}

The Westerbork Bootes Deep survey covers approximately 7 square
degrees and is centered at $14^{\rm h}32^{\rm m}05.75^{\rm s}$, $34^\circ
16' 47.5''$ (J2000). It consists of 42 discrete pointings, with
enough overlap to ensure a uniform sensitivity across the entire
field, with a limiting sensitivity of $28\mu$Jy
(1$\sigma_{\mbox{rms}}$). The catalog contains 3172 distinct
sources, of which 316 are resolved by the $13\times 27$ arcsec
beam. At these faint flux densities a significant fraction of the
sources originates in starbursting galaxies rather than AGN. One
of the aims of this survey is therefore to study the clustering
properties of starbursts. NOAO is carrying out a deep combined
optical and infrared study of this field which will provide
photometric redshifts for virtually all the radio sources. Since
these data are not yet available, Wilman et al. (2003)
\nocite{wil03} started the analysis of the clustering of the radio
sources with the study of the angular correlation function.
Applying the models of Overzier et al. (2003) to estimate what
fraction of measured amplitude of the correlation function was
found to be due to faint double FRII radio sources, led to the
conclusion that all the signal could be explained by these
sources.  
Upper limits on the cosmological clustering amplitude are,
however, consistent with the clustering of the radio-loud AGN being
diluted by the more weakly clustered IRAS-type starburst galaxies.

\subsection{Daddi-field}

With the initial aims of obtaining a well defined sample of
extremely red objects, Daddi et al. (2000) surveyed a 700 arcmin$^2$
region down to a limiting magnitude of approximately $K=19$.
\nocite{dad00a} and found about  400 objects with very 
red colours of $R-K > 5$. The subsequent analysis of the
clustering of these red objects found a large amplitude of the
correlation function of $>10^{-2}$. Furthermore, it appeared that
the amplitudes were a strong function  of $R-K$ colour in the range
$1<R-K< 7$, with the reddest objects being most strongly
clustered. These measurements were subsequently confirmed by other
groups using independent surveys (Roche et al. 2002, Firth et al.
2002). \nocite{roc02,fir02} This high signal is most easily
explained if the vast majority of these EROs are ellipticals in
the redshift range $1<z<1.5$. Such a narrow redshift range  stems
from a combination of two effects. First, at these magnitude
levels objects only are so red if the Balmer break is redshifted
beyond $z\sim 1$. Second, the bright magnitude level of K=19,
ensures that objects with $z>1.5$ are exceedingly rare in such
samples. Using such a narrow redshift distribution, the resulting
correlation length is 8-10 Mpc. While $r_0$ is large compared to
local field galaxies, it is consistent with predictions of
semi-analytic galaxy formation models, where the decrease in
clustering with redshift of dark matter halos is compensated by an
increase in their bias  (e.g. Kauffmann et al. 1999).
\nocite{kau99} This is further evidence that the majority of these
EROs are ellipticals rather than SCUBA type starburst galaxies.
Such starburst galaxies exhibit a large range in redshift, much
larger than expected for ellipticals in this sample. The resulting
correlation length inferred from the measured amplitude would then
be unrealistically large. This adds further weight that the
majority of EROs in this sample are $z\sim 1.5$ ellipticals.

\subsection{FIRES survey}

The Faint InfraRed Extragalactic Survey (FIRES) is a very deep
infrared survey centered on the Hubble Deep Field South using the
ISAAC instrument mounted on the VLT (Franx et al. 2000).
\nocite{fra00a} With integration times of more than 33 hours for
each of the infrared bands $J$, $H$ and $K$, limiting magnitudes of,
26.0, 24.9, and  24.5 respectivily are reached (Labb\'e et al. 2003).
\nocite{lab03}  A major advantage of observing this field is that
multicolour HST photometry is available resulting in accurate
photometric redshifts. A detailed analysis of the clustering of
galaxies in this survey was presented by Daddi et al. (2003).
\nocite{dad03} One of the interesting results was the finding of a
relatively large clustering amplitude for a $K$-selected sample down
to $K=24$, at a level comparable with measurements at $K\sim
19$. To explain this high level of clustering, the existence of a
highly clustered population of galaxies is required with a
comoving correlation length of $r\sim 10$ Mpc. To investigate the
clustering of distant galaxies further, we defined a sample of 105
objects with photometric redshift estimates $2<z<4$. The 48
galaxies with $J-K > 1.7$ have a correlation length of $8.2 \pm
1.1$ Mpc, a factor of about 2 larger than the bluer galaxies in
this sample. These red galaxies are a factor 3-4 more clustered
than Lyman break galaxies down to a level of $V_{606} = 27$. A
detailed analysis of the properties of this highly clustered
population, including characteristic mass and number density,
suggests that they are the objects that should be identified with
the progenitors of local massive early type galaxies.

\section{Clustering as a function of redshifts for galaxies and AGN}

The comparison of the clustering properties of galaxies and AGN as
a function of redshift is a powerful tool to constrain models of
their evolution. In addition to the measurements for clustering in
the samples just discussed, measurements for clustering of various
types of objects have been gathered from the literature (for
details see Overzier et al. 2003). \nocite{ove03}  For local
clusters, early and late type objects we took the measurements as
presented in Postman et al. (1992), Carlberg et al. (1997, 2000) 
and Norberg et al. (2002). \nocite{pos92} For the optical
quasars, we used the measurements for the 2dF quasars from Croom
et al. (2001). \nocite{cro01} For the very distant Universe, the
clustering analysis for Lyman break galaxies  are taken from Adelberger
(2000) \nocite{ade00} and Ouchi et al. (2001) \nocite{ouc01}

\begin{figure*}[t]
\centering
\centerline{\psfig{figure=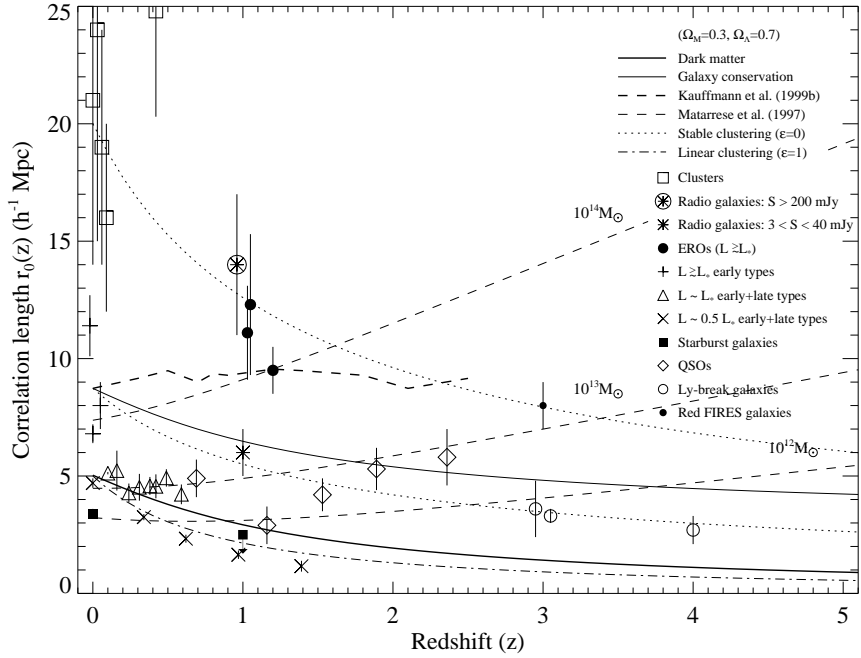,width=0.9\textwidth,clip=}}
\caption{\label{zoo} The redshift evolution of galaxy clustering
in a $\Lambda$CDM Universe. See Overzier et al. (2003) for
details. In addition to the plot presented by Overzier, also
plotted here are the correlation lengths $r_0$ for the FIRES galaxies
between $2 < z_{phot} < 4 $ from Daddi et al. (2003). A nice
representation of this figure showing actual images of the various
objects rather than symbols can be found at our website:
http://www.strw.leidenuniv.nl/\~{}overzier/r0.html.}
\end{figure*}

In Figure \ref{zoo} we show how the correlation length changes
as a function of redshift for the various galaxy and AGN samples.
In addition, a number of models for the evolution of clustering
are plotted.

The first important conclusion that can be drawn is that at
redshifts $z\simgt 1$ powerful radio sources are among the most
clustered population. Interestingly, the distant EROs have similar
correlation lengths. This suggest that EROs and powerful radio
galaxies might be similar objects at a different evolutionary
stage. This is supported by their K-band magnitudes being more or
less similar (e.g. Dunlop 2001). \nocite{dun01} Furthermore, the
number density of both classes is of the same order, provided that
a limited life time of the radio sources of the order of $10^7$
years is taken into account (Mohan et al. 2002; Willott et al.
2001). \nocite{wil01,moh02}

The inferred correlation lengths for the powerful radio sources is
a factor of 2-3 larger than that of the optical quasars. If this
is correct, then these optical quasars and radio sources can not
be ``unified'' through an evolutionary scheme. The much more
likely explanation is that QSOs in general reside in a field
population. Consistent
with many other lines of evidence (e.g. Best et al. 1998),
\nocite{bes98a} the powerful radio sources are hosted by very
massive systems.

In the local universe, clusters have the largest correlation
length. Detailed models of the evolution of clustering  (Refregier et
al. 2002), \nocite{ref02} indicate that $r_0$ of clusters should remain rather
constant with redshift. $z>1$ clusters and proto-clusters
therefore should be clustered at a level equal to or higher than
powerful radio galaxies. An obvious test whether there is a direct
link between powerful radio sources and clusters is to establish
that powerful radio sources are indeed often residing in clusters.

\section{Powerful radio galaxies as probes of distant clusters}

There a number of additional arguments that distant powerful radio
sources reside in (proto-)clusters (e.g. Best et al. 1998). An
important one is that these radio galaxies seem to reside in the
most massive, luminous and gas rich systems for a given redshift
and are therefore likely to be located in rich environments. For
example, the Hubble K-band diagram shows that at $z>1$ radio
galaxies are 1-2 magnitudes brighter than the brightest objects
found in ``field surveys'' (e.g. de Breuck et al. 2002).
\nocite{bre02} From dust, CO, HI absorption and emission line
studies, it is clear that radio galaxies contain a large reservoir
of gaseous material that is forming stars at high rates. (e.g.
Papadopoulos et al. 2000; \nocite{pap00} Archibald et al. 2000).
\nocite{arc00} The high star formation rates and deep spectroscopy
with 10-m class telescopes reveal UV spectra that are
remarkably similar to nearby star forming galaxies (Dey et al.
1997). \nocite{dey97}

Direct evidence that radio sources reside in dense environments
comes from the high rotation measures ($>1000$ rad / m$^2$) 
observed for 
a significant fraction of the $z>2$ radio sources (e.g. Carilli et
al. 1997). \nocite{car97a}  These are similar to the rotation
measures determined for radio sources that reside in the rich
local clusters.

The conclusive proof that at least a fraction of the radio sources
are in proto-cluster type environments comes from the discovery of
significant galaxy over-densities around a number of high redshift
radio galaxies. Currently the most efficient 
technique is to hunt
for Ly$\alpha$ emitting galaxies using a combination of
narrow-band imaging and multi object spectroscopy. The state of
the art in this field is presented in this workshop by Venemans.
In each of the 5 well-studied objects
with $2<z<4.1$  more than 20 
Ly$\alpha$ emitters have been found. Other techniques of
pinpointing galaxies are being developed. They rely on deep
mm/sub millimeter observations (e.g. Ivison et al. 2000),
\nocite{ivi00} sensitive X-ray imaging (e.g. Pentericci et al.
2002), \nocite{pen02} multi-colour imaging with one filter
blue-ward of the Lyman break, or infrared H$\alpha$ imaging. For
one proto-cluster (1138$-$262) at $z=2.2$ a number of these
techniques have been used, suggesting that the distribution of
H$\alpha$ emitters, Ly$\alpha$ emitters and extremely red objects
are all different (Kurk et al., these proceedings). This indicates that
those proto-clusters are comprised of a number of populations that
differ in age and/or metallicity.


\end{document}